# POLARIZED NEUTRON REFLECTOMETRY AT THE PRESENCE OF SMOOTH INTERFACIAL POTENTIAL

## Saeed S. Jahromi*, Seyed Farhad Masoudi

*Department of Physics, K.N. Toosi University of Technology, P.O. Box 15875-4416, Tehran, Iran;*

*E-mail: Saeed S. Jahromi, s.jahromi@dena.kntu.ac.ir*

## ABSTRACT

Polarized neutron reflectometry (PNR) have developed theoretically and experimentally in the past decades. In order to resolve the phase problem in neutron reflectometry, several simulation methods have been proposed such as reference layer and the variation of surroundings. By considering some factors such as smoothness or roughness in simulations, it is tried to gain more compatible results with experiment. In this paper, by using four different functions "Linear, Airy, Eckart and Error function", we investigate the effects of the smoothness of the interfacial potential on reflectivity, polarization of reflected neutrons and determination of the SLD of the sample, using reference layer method with polarized neutrons. We have also proposed a solution in order to gain better results with less noise in output reflectivity data, at the presence of smoothness.

## INTRODUCTION

Specular reflection of neutrons from nano-scale thin films can provide useful information about the physical and chemical properties of such layers and would help us to the study of surface structure [1]. By measuring the intensity of reflected neutrons, we can determine the scattering length density (SLD) of the sample along its depth. However the reconstruction of the surface profile has been hampered by the so-called phase problem. Like any scattering techniques in which only the intensities could be measured, in the absence of the phase, the least-squares fit methods could be used to determine the surface profile but generally more than one SLD could be find which corresponds with the same reflectivity data [1,2,4]. By knowing the phase, the real and imaginary parts of reflection coefficient, a unique result for the SLD profile could be retrieved with the help of the Gel'fan-d-Levitan integral equation [1,11-13].

In the past decades, several methods have proposed to find the phase of the reflection such as dwell time method and the reference layers which seems to be the best practical method [1]. The reference layer method which is based on transfer matrix method [2,7] was first achieved by Majkarzak et al. who also proposed and tested this method experimentally [1]. This method then developed by Leeb et al. by measurement of the polarization of reflected neutrons and a magnetic reference layer [3] and Masoudi et al. formulated the method in an straightforward way and enhanced the method [8-10].

In all of these works, it is supposed the interacting potential between neutrons and the sample at boundaries to be discontinuous and sharp. As we know from a realistic sample, there is some smearing at boundaries. Considering this smearing as a smoothness factor would cause some changes in the output reflectivity. In this paper by using four smooth varying functions; "Linear, Airy, Eckart and Error function", we have investigated the effects of the smoothness of interfacial potential on reflectivity, polarization of reflected neutrons and determination of the SLD of the sample, using reference layer methods with polarized neutrons. We have also proposed a solution in order to gain better results with less noise in output reflectivity data, at the presence of smoothness.

## THEORY

In neutron specular reflectometry one is dealing with a one dimensional scattering problem with potential V proportional to the SLD perpendicular to the surface. This scattering problem is represented by a one dimensional schrödinger equation:

$$\left(\partial_x^2 + (q^2 - 4\pi\rho(x))\right)\Psi(q,x) = 0 \qquad (1)$$

where $\rho(x)$ is the scattering length density of the sample and $q$ is the neutron wave vector in vacuum. The interacting potential between neutrons and the sample is represented by $v(x) = 2\pi\hbar^2\rho(x)/m$ [2].

For a magnetic layer, when the magnetization is in the plane of the film, the SLD is defined by $(\rho_\pm = \rho(x) \pm \mu \cdot B)$ where $\rho(x)$ is the depth profile of none magnetized film, $\mu$ is the magnetic moment of the incident neutrons and B is the magnetic field due to the magnetization of the film. The plus and minus sign is proportional to the polarization of incident neutrons, parallel and anti parallel to the local magnetization, respectively [2,3]. Here we suppose the magnetic induction outside the ferromagnetic is small enough not to affect the neutron beam.

An alternative exact representation of reflection coefficient for such a layer, $r_\pm(q)$, is derived from the transfer matrix method of solving eq (1). The elements of the 2×2 transfer matrix, $A(q), B(q), C(q), D(q)$ for arbitrary surround can be expressed as [1,2]:

$$\begin{pmatrix} 1 \\ ih \end{pmatrix} t_\pm e^{ihqL} = \begin{pmatrix} A & B \\ C & D \end{pmatrix} \begin{pmatrix} 1+r_\pm \\ i(1-r_\pm) \end{pmatrix} \quad (2)$$

where $h = (1 - 4\pi\rho_s / q^2)^{1/2}$ is the refractive index of the substrate with SLD $\rho_s$.

The reflectivity, $R_\pm(q) = |r_\pm(q)|^2$, can be related to the elements of the transfer matrix in term of new quantity $\Sigma_\pm(q)$ [1,4]:

$$\Sigma_\pm(q) = 2\frac{1+R_\pm}{1-R_\pm} = h(A^2 + B^2) + \frac{1}{h}(C^2 + D^2) \quad (3)$$

The reflection coefficient in term of three new parameters, $\gamma^h, \beta^h, \alpha^h$, can be expressed as:

$$r(q) = \frac{\beta^h - \alpha^h - 2i\gamma^h}{\beta^h + \alpha^h + 2} \quad (4)$$

where

$$\alpha^h = hA^2 + h^{-1}C^2$$
$$\beta^h = h B^2 + h^{-1}D^2 \quad (5)$$
$$\gamma^h = h AB + h^{-1}CD$$

*Reference method*

Suppose the sample is separated into two distinct regions. An unknown layer and the known magnetized reference film mounted between the substrate and the unknown film.

The total transfer matrix for such a sample is represented as [7]:

$$\begin{pmatrix} A & B \\ C & D \end{pmatrix} = \begin{pmatrix} w_\pm & x_\pm \\ y_\pm & z_\pm \end{pmatrix} \begin{pmatrix} a & b \\ c & d \end{pmatrix} \quad (6)$$

where $(w_\pm,...,z_\pm)$ is transfer matrix of the known part and $(a,...,d)$ describes the transfer matrix correspond to the unknown part. (+) and (−) denote the plus and minus magnetization with respect to the polarization of incident neutrons [3].

By using eq (6), the parameter $\Sigma_\pm(q)$ can be expressed as [8]:

$$\Sigma_\pm(q) = \beta^h_{k\pm}\tilde\alpha_u + \alpha^h_{k\pm}\tilde\beta_u + 2\gamma^h_{k\pm}\tilde\gamma_u \quad (7)$$

where the tilde represents the reversed unknown film; that is, the interchange of the diagonal elements of the corresponding transfer matrix $(a \leftrightarrow d)$ [1].

To derive the reflection coefficient of the sample as a function of polarization of reflected neutrons, we suppose the incident neutrons to be polarized in the direction normal to the reflection surface ($x$) and the reference layer is magnetized in a direction parallel to the reflecting surface ($z$). In this case the polarization of reflected neutrons is represented as follows [3]:

$$P_x + iP_y = \frac{2r_+^* r_-}{R_+ + R_-}$$
$$P_z = \frac{R_+ - R_-}{R_+ + R_-} \quad (8)$$

By using eq (5) to (8) for the polarization of reflected neutrons, we can write [8-10]:

$$P_x = 1 + \frac{2\zeta}{\Sigma_+ \Sigma_- - 4} \quad (9)$$

$$P_y = \frac{2\zeta}{\Sigma_+ \Sigma_- - 4}(c_{\gamma\beta}\tilde\alpha_u + c_{\alpha\gamma}\tilde\beta_u + c_{\alpha\beta}\tilde\gamma_u) \quad (10)$$

$$P_z = 2\frac{\Sigma_+ - \Sigma_-}{\Sigma_+ \Sigma_- - 4} \quad (11)$$

where

$$\zeta = 2(1 + \gamma^h_{k+}\gamma^h_{k-}) - (\alpha^h_{k+}\beta^h_{k-} + \beta^h_{k+}\alpha^h_{k-}) \quad (12)$$

and

$$c_{ij} = i^h_{k+} j^h_{k-} - j^h_{k+} i^h_{k-} \quad (13)$$

for $i$ and $j$ = '$\alpha$', '$\beta$' and '$\gamma$'. The parameters $c_{ij}$ and $\zeta$ are known from the elements of the transfer matrix of the know reference layer.

By knowing $\Sigma_+$, $\Sigma_-$ and $P_y$, we can determine the parameters $\gamma^h, \beta^h, \alpha^h$ for the unknown part of the sample, $\Sigma_\pm$ is determined directly from measurement or by using equation (9) to (13) as follows [8,10]:

$$\Sigma_\pm^2 \pm \frac{\zeta P_z}{2(P_x - 1)}\Sigma_\pm - (4 + \frac{\zeta}{(P_x - 1)}) = 0 \quad (14)$$

Eq (14) has two different solutions. The physical solution is selected from the fact that $\Sigma_\pm > 2$ [8]. Actually by knowing two of the parameters, $\Sigma_+$, $\Sigma_-$, $P_x$ and $P_z$, the two others could be determined.

As an example, by using $P_y$, $P_z$ and $\Sigma_-$, the three parameters are calculated as follows:

$$\begin{pmatrix} \tilde\alpha_u \\ \tilde\beta_u \\ \tilde\gamma_u \end{pmatrix} = M^{-1} \begin{pmatrix} \Sigma_- \\ \dfrac{4P_z\Sigma_- - 2\Sigma_-}{P_z\Sigma_- - 2} \\ \dfrac{2P_y P_z\Sigma_- - P_y\Sigma_-^2}{P_z\Sigma_- - 2} - 2P_y \end{pmatrix} \quad (15)$$

$$M = \begin{pmatrix} \beta_{k-}^h & \alpha_{k-}^h & 2\gamma_{k-}^h \\ \beta_{k+}^h & \alpha_{k+}^h & 2\gamma_{k+}^h \\ c_{\gamma\beta} & c_{\alpha\gamma} & c_{\alpha\beta} \end{pmatrix} \qquad (16)$$

The parameters $\tilde{\alpha}_u, \tilde{\beta}_u$ and $\tilde{\gamma}_u$ denote the free reversed unknown sample.

### Smooth interfacial potentials

As we mentioned in chapter 1, we are going to investigate the effects of the smoothness of interfacial potential at boundaries by choosing four continuous and smooth varying functions; 'Linear, Airy, Eckart and Error function'.

### Linear potential

In this method the SLD of two adjacent layers with the thicknesses $d_1$ and $d_2$, is supposed to vary linearly from the value $\rho_1$ to $\rho_2$ at the interface. By considering the linear variation area to be a distinct layer with the thickness of $\varepsilon(d_1 + d_2)$, where $\varepsilon$ is a parameter called smoothness factor which is $0 \le \varepsilon \le 1$, our sample is divided into three region; two rectangular film with the thicknesses $d_1(1 - \varepsilon)$, and $d_2 (1 - \varepsilon)$ and SLD $\rho_1$ and $\rho_2$ respectively and the linear film. As the parameter $\varepsilon$ is increased, the thickness of the linear layer increased too, and the interfacial potential is supposed to be smoother. $\varepsilon=0$ denotes the non-smooth potential.

The total transfer matrix for the entire sample is derived from the multiplication of the transfer matrix of the three parts. In order to determine the transfer matrix of the linear part, we can use one of the following methods; first, the step method and second, the direct solution of the schrödinger for the linear potential which is the Airy function.

In step method, we divide the linear part into several small rectangular layers and multiply the transfer matrix of individual films, in this manner, the SLD of each layer increase or decrease like a step depending on the gradient of the linear function for plus and minus gradient, respectively.

The direct solution of the schrödinger equation, Eq (2), for the linear potential of previous chapter, with the smoothness factor $\varepsilon$, is presented as:

$$\psi(x) = C^+ Ai(\chi) + C^- Bi(\chi) \qquad (17)$$

where $Ai$ and $Bi$ are the Airy function of the first and second kind, respectively and;

$$\chi = \frac{\eta x - \lambda}{(\eta)^{2/3}}$$

$$\eta = 4\pi m$$

$$\lambda = q^2 - 4\pi\rho_1$$

$$m = \frac{(\rho_2 - \rho_1)}{\varepsilon(d_2 + d_1)} \qquad (18)$$

The elements of the transfer matrix for the linear layer are determined by using boundary conditions at the interfaces.

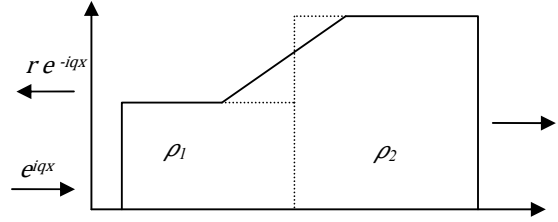

**Figure 1.** *The linear potential at the interface of two adjacent layer.*

As shown in fig 1, for an incoming from the left, we have:

$$A = \frac{Bi'(\chi)|_{x=x_0} Ai(\chi)|_{x=x_L} - Bi(\chi)|_{x=x_L} Ai'(\chi)|_{x=x_0}}{Ai(\chi)|_{x=x_0} Bi'(\chi)|_{x=x_0} - Bi(\chi)|_{x=x_0} Ai'(\chi)|_{x=x_0}}$$

(17-a)

$$B = \frac{q}{\lambda^{1/3}} \times \frac{Ai(\chi)|_{x=x_0} Bi(\chi)|_{x=x_L} - Bi(\chi)|_{x=x_0} Ai(\chi)|_{x=x_L}}{Ai(\chi)|_{x=x_0} Bi'(\chi)|_{x=x_0} - Bi(\chi)|_{x=x_0} Ai'(\chi)|_{x=x_0}}$$

(17-b)

$$C = \frac{\lambda^{1/3}}{q} \frac{Ai'(\chi)|_{x=x_L} Bi'(\chi)|_{x=x_0} - Ai'(\chi)|_{x=x_0} Bi'(\chi)|_{x=x_L}}{Ai(\chi)|_{x=x_0} Bi'(\chi)|_{x=x_0} - Bi(\chi)|_{x=x_0} Ai'(\chi)|_{x=x_0}}$$

(17-c)

$$D = \frac{Ai(\chi)|_{x=x_0} Bi'(\chi)|_{x=x_L} - Bi(\chi)|_{x=x_0} Ai'(\chi)|_{x=x_L}}{Ai(\chi)|_{x=x_0} Bi'(\chi)|_{x=x_0} - Bi(\chi)|_{x=x_0} Ai'(\chi)|_{x=x_0}}$$

(17-d)

By knowing the total transfer matrix, the reflectivity for the whole sample is known.

### Error function

In this method, the smooth variation of the SLD of two adjacent films in term of layers depth is represented by [6]:

$$\rho(x) = \rho_1(x) + \frac{\rho_2(x) - \rho_1(x)}{2}[1 + erf(\frac{x - \Delta}{\sqrt{2}\,\sigma})] \qquad (18)$$

where $\sigma$ is a parameter, which denotes the smoothness and $\Delta$ is the turning point of the error function.

### Eckart potential

The smooth variation of the potential at boundaries can also be expressed by the so-called Eckart potential which is one of the most applicable functions in nuclear physics. The SLD variation corresponding with the Eckart potential is as follows [5]:

$$\rho(x) = \rho_1(x) + [\rho_2(x) - \rho_1(x)]/[1 + \exp(\frac{x - \Delta}{b})] \qquad (19)$$

where $b$ is the smoothness factor and $\Delta$ is the turning point of the Eckart potential.

# EXAMPLE

To investigate the effects of the smoothness of the potential on reflectivity, polarization of reflected neutrons and the SLD, we consider a two layer sample composed of 20 nm thick Copper over a 15 nm thick gold with the SLD of 6.52 and 4.46 ×$10^{-4}$ $nm^{-2}$ for Copper and gold, respectively and a magnetic 20 nm Cobalt film with the SLD of 6.44 and -1.98 ×$10^{-4}$ $nm^{-2}$ for plus and minus magnetization, respectively as reference layer, which is mounted between the sample and a silicon substrate. It is supposed the incident neutron beam is polarized in the $x$ direction and the magnetic field of the Cobalt reference is along the $z$ axis.

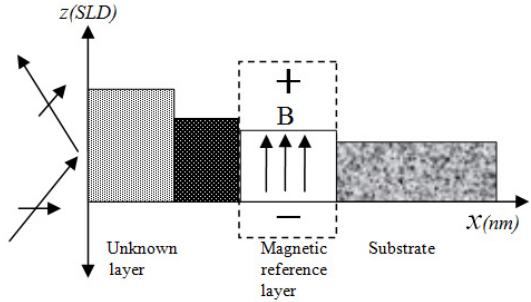

*Figure 2. Arrangement of a sample for investigating the smoothness of the potential. The dashed line represents the effective potential experienced by neutrons parallel and anti parallel to the magnetic field **B***

Figure 3-a, shows the polarization of the reflected neutrons form the sample of figure 2 for the Eckart potential with smoothness factor $b=5A^o$. The real and imaginary parts of the reflection coefficient for the Error function with $\sigma=5A^o$, is illustrated in Figure 3-b. The reflectivity and phase of the reflection for the linear interfacial potential is depicted in Figure 4, **a** and **b**, using step method and airy function with smoothness factor $\varepsilon=0.05$ and 0.02, respectively.

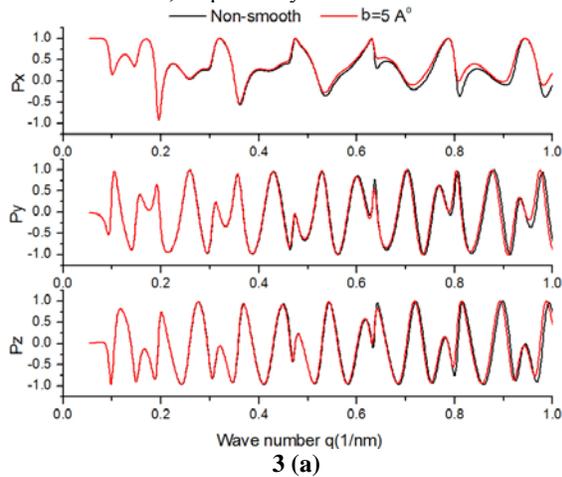

3 (a)

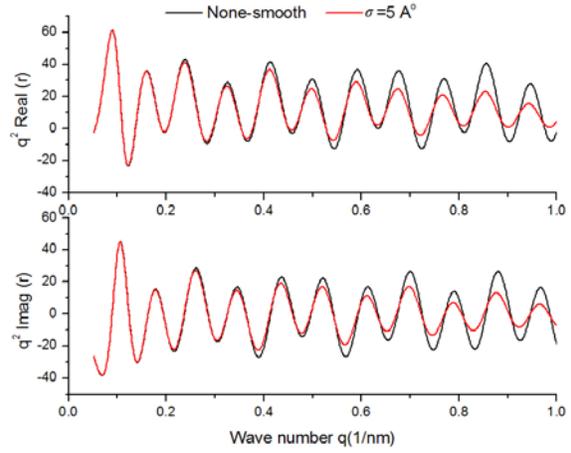

3 (b)

*Figure 3. **a**) Polarization of the reflected neutrons for the Eckart potential with smoothness factor; $b=5A^o$. **b**) Real and imaginary parts of the reflection coefficient for the Error function with smoothness factor $\sigma=5A^o$.*

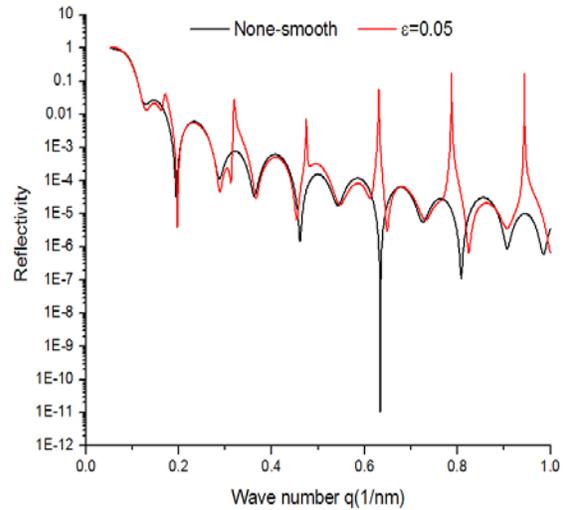

4 (a)

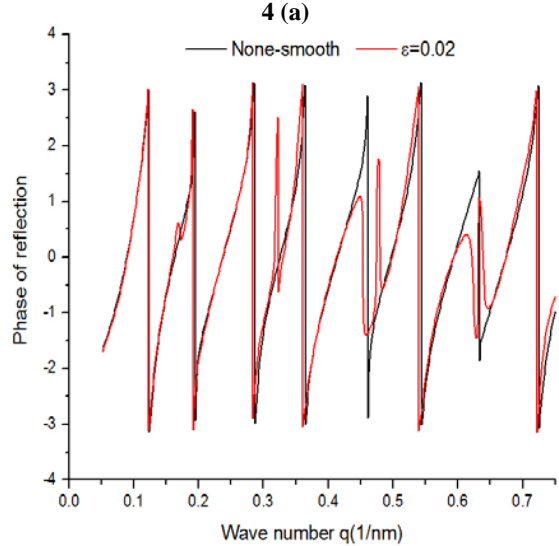

4 (b)

***Figure 4.** **a)** Reflectivity curve for the linear potential (step method) with $\varepsilon=0.05$. **b)** Phase of the reflection for the linear potential (Airy function) with $\varepsilon=0.02$.*

As it is presented in Figure 3 and 4, the effects of the smoothness of the interfacial potential on output results, is clear in all of the figures specially at large wave numbers. The results for the small wave numbers, $q \leq 0.6$, truly correspond with none-smooth data.

Figure 4-a,b demonstrate some noises in reflectivity and phase of the reflection even in small wave numbers for the linear potential. These noises are due to the abrupt changes in the elements of the transfer matrix of the whole sample, M (Eq. (16)). As the noises are existed even on the real and imaginary parts of the reflection coefficient graphs, we can't use them to reconstruct the SLD of the sample.

As the total transfer matrix depends on the reference layers, we can diminish or eliminate the noises or shift them to the $q$ value larger than 0.6, by finding a suitable thickness for the reference layers. The rest of the noises can be removed by extrapolation.

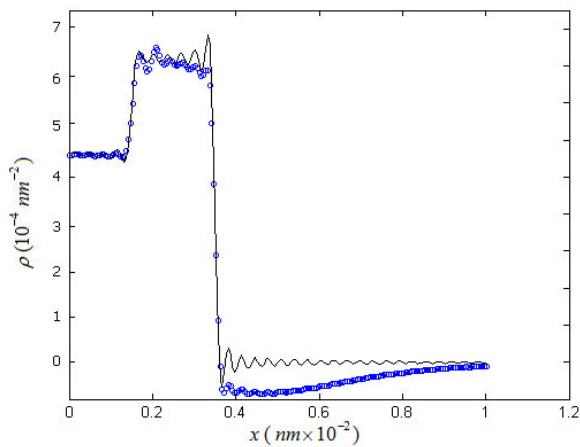

**5 (a)**

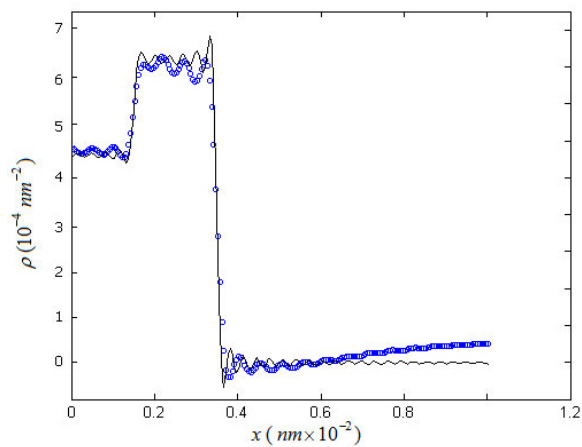

**5 (b)**

***Figure 5.** Reconstructed SLD of the free reversed unknown sample; **a)** Error function, **b)** Linear potential (step method). The blue circled curves demonstrate the reconstructed SLD of the sample of Figure 2 at the presence of smoothness.*

In order to retrieve the SLD of the unknown sample, the data for the real and imaginary parts of the reflection coefficient are needed as input data for some useful codes like the one which is developed by P. Sacks [11] based on Gel'fan-d-Levitan integral equation [11-13].

To show the stability of the reconstruction process of the SLD at the presence of the smoothness, the SLD of the unknown sample was retrieved for the Error function and linear potential (step method). As there were no noises in the data of Error function, they could be used directly as input data for Sacks code, however the data of linear potential needed to get clear from noises. The best result with less noises, obtained for the thickness of 10 nm for Cobalt reference. In order to remove the noises at large wave numbers, the data of $q$'s larger than 0.6 was also neglected. Then these purified data was used as input for Sacks code.

The retrieved SLD of the free reversed unknown sample is illustrated in Figure 5-**a** and **b** for the Error function and linear potential, respectively. The 15 nm thick Copper with the SLD of 6.52 and 20 nm thick gold with SLD of 4.46, is clearly demonstrated in the Figure.

## CONCLUSION

In most of simulation method in PNR, it is supposed the potential of two adjacent layers to be discontinuous and sharp. In this paper by introducing the reference method for retrieving the phase of the reflection for polarized neutrons, we investigated the effects of the smoothness of the interfacial potential on, reflectivity, polarization of reflected neutrons and the SLD of the sample for four smooth varying functions (Linear, Airy, Eckart and Error function).

The output results show that; depending on the range of the neutron wave numbers, consideration of the smoothness is very important. The output data for small wave numbers is clearly correspond with non-smooth data, nevertheless at large values of $q$, the data deflects from the none-smooth curve.

We can decrease the effects of smoothness on output data by selecting a suitable thickness for the reference layers and using the data of wave numbers smaller than 0.6.

It was shown that the reconstruction process of the SLD is yet stable at the presence of the smoothness.